\documentclass[10pt,conference]{IEEEtran}
\IEEEoverridecommandlockouts
\usepackage[numbers]{natbib}
\usepackage{amsmath,amssymb,amsfonts}
\usepackage{algorithmic}
\usepackage{graphicx}
\usepackage{textcomp}

\def\BibTeX{{\rm B\kern-.05em{\sc i\kern-.025em b}\kern-.08em
		T\kern-.1667em\lower.7ex\hbox{E}\kern-.125emX}}

\newcommand{\RNum}[1]{\uppercase\expandafter{\romannumeral #1\relax}}
\usepackage{ragged2e}
\usepackage{array}
\usepackage{subfig}
\usepackage[table]{xcolor}
\definecolor{mygray}{gray}{0.9}
\usepackage{multirow}
\usepackage{float}
\DeclareCaptionFont{mysize}{\fontsize{8}{9.6}\selectfont}
\usepackage{dblfloatfix}
\newcommand{\comment}[1]{}
\begin{document}
	
	\title{Deep Neural Network Approach to Estimate Early Worst-Case Execution Time
	}
	
	\author{\IEEEauthorblockN{ Vikash Kumar}
		\IEEEauthorblockA{\textit{Computational and Data Sciences} \\
		\textit{Indian Institute of Science}\\
		Bangalore, India\\
		kvikash@iisc.ac.in}

}

	\maketitle
	
	\begin{abstract}
		Estimating Worst-Case Execution Time (WCET) is of utmost importance for developing Cyber-Physical and Safety-Critical Systems. The system's scheduler uses the estimated WCET to schedule each task of these systems, and failure may lead to catastrophic events.  It is thus imperative to build provably reliable systems. WCET is available to us in the last stage of systems development when the hardware is available and the application code is compiled on it. Different methodologies measure the WCET, but none of them give early insights on WCET, which is crucial for system development. If the system designers overestimate WCET in the early stage, then it would lead to the overqualified system, which will increase the cost of the final product, and if they underestimate WCET in the early-stage, then it would lead to financial loss as the system would not perform as expected.
		
		This paper estimates early WCET using Deep Neural Networks as an approximate predictor model for hardware architecture and compiler. This model predicts the WCET based on the source code without compiling and running on the hardware architecture. Our WCET prediction model is created using the Pytorch framework. The resulting WCET is too erroneous to be used as an upper bound on the WCET. However, getting these results in the early stages of system development is an essential prerequisite for the system's dimensioning and configuration of the hardware setup.
	\end{abstract}
	
	
	\section{{Introduction}}
	In safety-critical systems, the timing domain is as important as the value domain. These systems need to satisfy the timing constraint; otherwise, resource damage or even life loss could occur. For instance, it is essential to know that airbags in cars open fast enough to save lives. Besides, these systems not only satisfy the correctness of the system but also be responsive. If the system does not satisfy the timing constraints after manufacture, then changing the hardware that cannot schedule tasks would be more expensive to redesign. Therefore, estimating the worst-case execution time is very crucial. Estimating WCET \cite{wilhelm2008worst} for the given architecture is difficult, if not impossible, to cover all the system states, and it requires the user's input. Modern processors are equipped with complex architectural features such as superscalar pipelines and caches that make WCET estimation complex. For instance, caches introduce the variance in operations execution time based on the hit or miss in the caches. In the previous decade, many optimizations have been done to improve the average-case execution time, but less work has been done to estimate WCET precisely and accurately. 
	
	The process of estimating the WCET is called timing analysis. The timing analysis of the given system is possible in the last stage of the system development process. Hardware architecture and compiled binary code are required to estimate the WCET. By getting the early estimate of WCET, we will prune the system's unwanted design points based on the parameter of interest, i.e., design system exploration. There are three popular ways of estimating WCET, i.e., the Measurement-based approach, Static Analysis, and the Hybrid approach. The Measurement-based method \cite{wenzel2005measurement} executes the task on the given hardware or the simulator for the different inputs and states (initial and intermediate) of the architecture to measure the execution time. All different sets of inputs data are applied to measure the maximal program execution time. The measurement-based approach is the most common technique in the industry because hardware and simulators are usually available. The main disadvantage of this method is that determining the inputs to be considered for the WCET is not obvious, and running the WCET analysis over the entire set of possible inputs is not feasible.
	
	On the other hand, the Static approach \cite{heckmann2004worst} does not execute the task on hardware or simulator but analyzes the set of possible control flow and reduces the number of different possible inputs using safe abstraction. The success of the static approach is exposed by vendors of the hardware, but in recent years, hardware vendors do not reveal their system features anymore. This approach also has the same shortcomings as the measurement-based approach, in that it overestimates results due to the lack of information. Additionally, the complexity of the static approach leads to increased code size, which is not preferable.
	
	In the Hybrid approach \cite{petters2000bounding}, we combine the concepts from the Measurement-based and Static approach. The hybrid approach identifies a single feasible path, a program path consisting of a basic block sequence. The advantage of this approach is that it does not rely on the abstract model of the hardware architecture. However, instrumented code is required, which may not be allowed in all cases, and correct WCET is not possible because safe initial state and worst-case input can not be assumed. WCET analysis tools such as aiT \cite{ferdinand2004ait}, Chronos \cite{li2007chronos}, Heptane \cite{hardy2017heptane}, and OTAWA \cite{casse2006otawa} are used to determine the safe upper bound when the hardware architecture and compiled binary code are available. All the existing WCET analysis tools are incapable of predicting early WCET in the system's development process. To overcome this problem, we are proposing an approach to predict early WCET without using the binaries.
	
	In this paper, we have extended the work done in \cite{altenbernd2016early}. The linear model presented in \cite{altenbernd2016early} will no longer give better accuracy as the system's complexity increases, such as pipeline, caches, and branch prediction. We employ Deep Neural Networks to predict early WCET instead of running the application source code physically on the hardware. We use a Measurement-based approach among various WCET estimation strategies available. To our best knowledge, nobody uses the Deep Neural Network model to estimate early WCET on the available datasets.
	
	The remaining of the paper is structured as follows: Section \RNum{2} presents related work, and in section \RNum{3}, we brief Deep Neural Networks. Section \RNum{4} describes our WCET estimation approach to predict early WCET using neural networks, which is evaluated in section \RNum{5}. Experimental results are reported in section \RNum{6}. We conclude the paper finally and present the prospects of future work in section \RNum{7}.

		\section{{Related work}}
		Bonenfant et al. \cite{bonenfant2017early} presented an approach for early WCET prediction using machine learning based on C source code. Their method used a Static approach which generated worst-case event counts such as the number of arithmetic operations like addition, subtraction, multiplication, and division, the number of function calls, the number of global variables, and the number of reads and writes access. To train the model, they used these features with labeled WCET. The worst-case events count of source code was formulated to obtain a satisfiable prediction of the future WCET. As far as considering estimating early WCET, this approach works well. However, it has some shortcomings in that event-counting of code using a Control Flow Graph (CFG) results in the loss of valuable code flow information.
		
		Thomas Huybrechts et al. \cite{huybrechts2018new} proposed a new extension to the hybrid approach to predict early WCET using machine learning. This new approach estimates the WCET using smaller entities of the code, so-called hybrid blocks, based on software and hardware features. As a result, the ML-based hybrid analysis provides insight into the WCET early on in the development process and refines its estimate when more detailed features are available. A new tool named COBRA was proposed to extract the features. The extracted features were used to train and validate the model. Machine learning approaches, such as Linear regression, Tree regression, and Support vector machines, were used to compare the results of TACLeBench \cite{falk2016taclebench} applications. The mean relative error for support vector regression with Linear kernel was 40.2\%, which was too high to use as an upper bound on WCET.
		
		Oyamada et al. \cite{oyamada2004accurate} presented a neural network-based approach for accurate software performance estimation, which also deals with the non-linear behavior of execution times due to complex modern architecture such as deep pipeline, branch prediction, and cache sizes. Assembly instructions were used as features categorized as floating-point, integers, branches, and load/store operations. Based on the CFG, the trained data is classified into two parts as control dominated applications and data dominated applications. Feed-Forward neural networks have been used with one input layer, one hidden layer, and one output layer with different neurons at each layer. CFG weights were used to make the distinction of application domains. The generic estimator had a maximum overestimation of 41.01\%  and a maximum underestimation of 20.69\%. However, for the specialized estimator, they improved the overestimation and underestimation slightly. The error was too high for the estimate to be used as upper bound but obtaining such results in the development process is useful for system design.
		
		The approach proposed in \cite{huybrechts2018introduction} extended the work done in \cite{huybrechts2018new} with a deep neural network to estimate WCET. This work used two different models: a feed-forward neural network and a tree recursive neural network. The data used in their experiments was taken from TACLeBench  benchmark suits. The architecture used for dataset A was one input layer of ten neurons, three hidden layers, 32 neurons, and one output layer. The results were given in terms of Root Mean Square Error (RMSE), and for dataset A, it was around 40\% on validation. The results are too large  to obtain any useful upper bound. 
		
		Lisper and Santos \cite{lisper2009model} developed a new Measurement-based WCET analysis method, which uses regression to identify parameters in the common linear Implicit Path Enumeration Technique (IPET) model to calculate WCET. The method can use different granularity timing measurements, including end-to-end measurements, which reduces the need for fine-grained timing measurement instrumentation.
		
		Abel and Reineke \cite{Abel13measurement-basedmodeling} developed an algorithm to model the cache's replacement policy by measuring actual hardware automatically. This work helps identify the cache-sensitive timing model.
		
		This paper presents a Deep Neural Network based approach to predict early WCET with a network architecture different from the aforementioned approaches. Our models are evaluated on TACLeBench \cite{falk2016taclebench} benchmark suites. We use the SWEET \cite {lisper2014sweet} tool to extract the features. The primary function of SWEET is to perform \textit{flow analysis} to identify \textit{flow facts} i.e., information about loop bounds and infeasible paths in the program. Flow facts are necessary for finding safe and tight WCET. Any WCET analysis must satisfy safeness and tightness conditions, which reflects the estimate of WCET precisely.
\\ \\

		\section{{Deep Neural Network}}
		Deep Neural Networks \cite{lecun2015deep} are gaining popularity in every field of life due to their ability to solve complicated applications with increasing accuracy over time. They are a subset of Artificial intelligence that attempt to learn patterns based on input data.
		It is the machine learning techniques that provide computers with the ability to learn from observed data. Supervised learning and Unsupervised learning are different types of machine learning. In supervised learning, the system is given labeled data, whereas in unsupervised learning, the system is given unlabeled data. Our approach uses supervised learning in which labels are formed out of the number of cycles consumed for each training program. This will be further explained in section V.
		
\begin{figure}[b]
			\centerline{\includegraphics[width=3.5in]{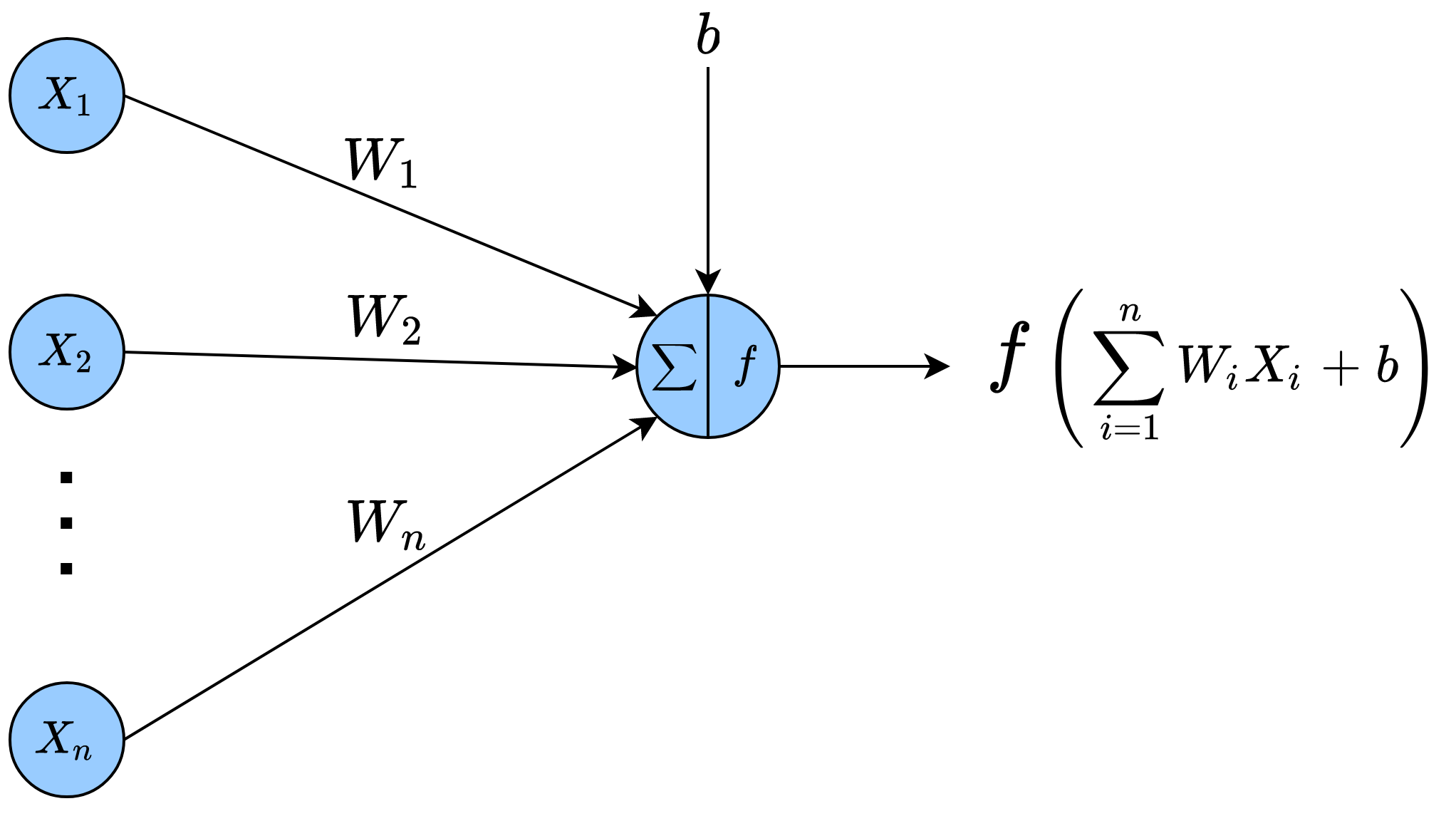}}
			\caption{A Neuron.}
			\label{fig 1}
			\end{figure}

		The analogy of a neural network \cite{lecun2015deep} has been taken from the neurons present in the human brain. The whole concept of deep learning is to try and mimic the human brain and get similar functions as the human brain has and leverage the things that evolution has already developed for us. 
		Millions of neurons are present in the human brain. Neurons send and process signals in the form of electrical and chemical signals. Biological neural networks consist of interconnected neurons with dendrites that receive inputs. Based on these inputs, they produce an output through an axon to another neuron. The neuron (node) is the building block of any deep neural network. An example of a neuron in fig.~\ref{fig 1}. shows the input (X\textsubscript{1} - X\textsubscript{n}), their corresponding weights (W\textsubscript{1} - W\textsubscript{n}), a bias (b) and the activation function ƒ applied to the weighted sum of the inputs. The parts/components of a typical deep learning system are described below, and later we apply the following steps to create the model, train the model, and for the accuracy of the model.  
		
			\begin{figure}[htbp]
			\centerline{\includegraphics[width=3.5in]{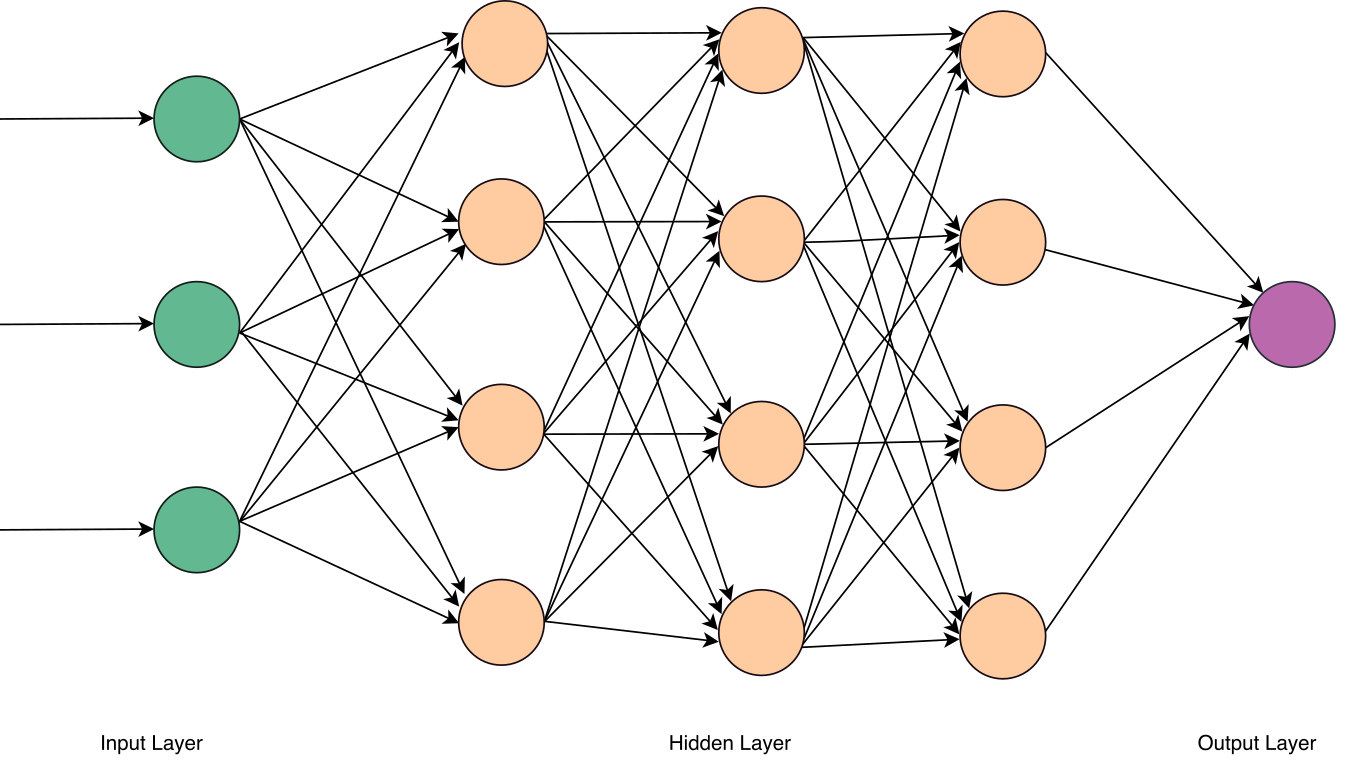}}
			\caption{A Feed Forward Neural Network.}
			\label{fig 2}
			\end{figure}
		
		\begin{itemize}
			\item Data - The data is what we apply deep learning techniques on. The data gives insights into how our features and labels are related.
			\item Task - On the given data, what tasks have to be performed -- such as classification and regression.
			\item Model - Model represents the details of the architecture. Some of the popular models are Feed-Forward Neural Network, Convolution Neural Network, and Recurrent Neural Network. 
			\item Loss Function - The loss function evaluates how well the learning algorithm predicts the outcome. The learning algorithm tries to improve itself by loss functions. There are different types of loss functions such as mean square error and cross-entropy loss.
			\item Learning Algorithm - The learning algorithm is used to update each parameter in our neural network. Using a learning algorithm, our model learns to identify trends in the data. Some of the different learning algorithms are gradient descent and Adam \cite{kingma2014adam}.
			\item Accuracy - The predicted value is compared with the actual value to find the accuracy which tells us how well our network performs.
		\end{itemize}
		
			\begin{figure*}[htbp]
			\centerline{\includegraphics[width=7in]{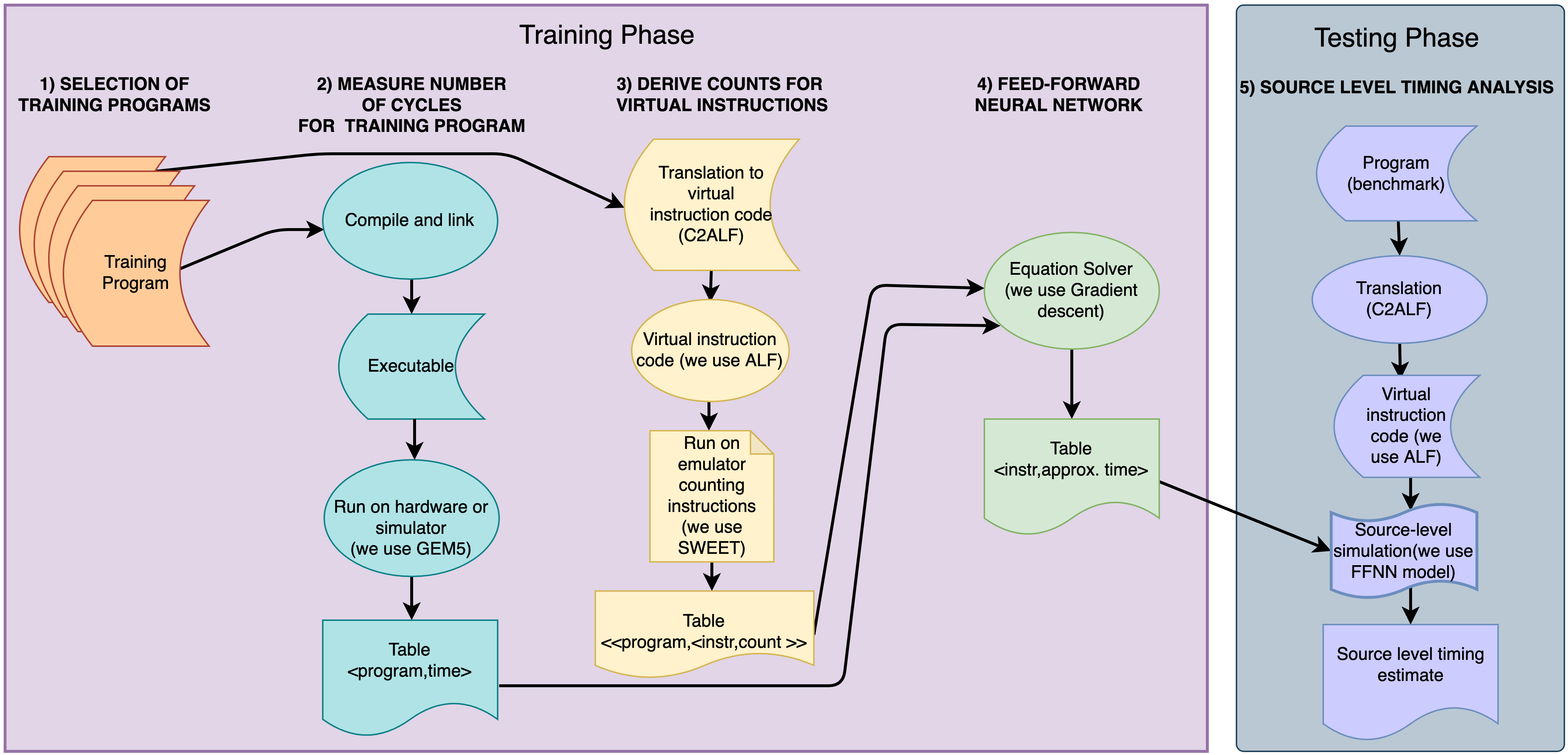}}
			\caption{Early-timing analysis approach.}
			\label{fig 8}
	    	\end{figure*}
    	
    		\begin{table*}[htbp]
    		\renewcommand{\arraystretch}{1.5}
    		\caption{FEATURES EXTRACTED}
    		\label{table_example1}
    		\centering
    		\begin{tabular}{|l|l|l|}
    			\hline
    			Number of addition operations&Number of subtraction operations&	Number of multiplication operations\\
    			\hline
    			Number of division operations &Number of logic operations&Number of shift operations\\ 
    			\hline
    			Number of function calls&Number of return statements&Number of jump statements\\
    			\hline
    			Number of load operations& Number of store operations&Number of comparison operations\\
    			\hline
    		\end{tabular}
    	\end{table*}
	
		A feed-forward neural network (FFNN) or multilayer perceptron (MLP) \cite{svozil1997introduction} is an essential deep learning model. A feed-forward network, as shown in fig. 2. consists of one input layer, one and more hidden layers, and one output layer. In the feed-forward network, neurons are not connected to themselves or neurons in the same layer. A fully connected network is a particular case of a feed-forward network where all neurons of one layer are connected to all the following layer's neurons. In the general case, not all the neurons need to be active, and in some networks, most of them are inactive. Neurons at the hidden layer have two portions -- linear, and non-linear activation functions. For the given inputs and weights of each layer, the feed-forward network can predict the desired output. This process is known as a forward pass in deep learning terms. Later, using backpropagation, we update our models' parameters.

	\section{Early WCET Estimation using DNN}
	The detailed explanation of the proposed approach is explained below. Figure 3 depicts both the training and testing phases.
	
	\begin{itemize}
		\item A selection of training programs is made.
		\item A training program is compiled and executed on the target hardware or a simulator. The execution time is measured for each program.
		\item The training program is converted into a virtual instruction set. With the help of SWEET tool, the instruction count of the virtual instructions are recorded.
		\item The data is fed to the feed-forward neural network. In contrast to work done in [9], this step of WCET estimation is entirely different as they have used a linear model in this step, and this approach uses a Feed-Forward Neural Network.
		\item Finally, the derived feed-forward neural networks model is used for timing analysis.
	\end{itemize}

\section{Experimental Setup}
	Selection and construction of training programs are of utmost importance. Each training program is constructed using the extended approach presented in \cite{altenbernd2016early}, which covers all the context-dependent timing effects due to hardware features such as caches, pipelines and branch prediction units, and code optimizations due to the compiler. The training programs are executed on the gem5 \cite{binkert2011gem5} simulator to measure the number of cycles which are used as labels. The same training programs are translated into a virtual instruction set using SWEET \cite {lisper2014sweet} tool. This virtual Instruction Set Architecture (ISA) acts as a feature set of the proposed predictor network. Combining these two essentially creates the data which can be appropriately used by neural networks.
	
	In deep learning, we need to pre-process and clean the data before feeding it to a neural network. We have used feature selection on our training data, and we found that the features shown in table~\ref{table_example1} are the most frequently occurring out of all the features.  So we need to choose the features carefully  \cite{guyon2003introduction}. We found that some features have values in the range of 200 - 800, and some features have values in the range of 2 - 10. These differences in the range of values bias the model's prediction to be inclined towards values of features that are larger, and the features having lower values contribute much less to the prediction, i.e., low value features have no significance in the Neural Network \cite{hall2000correlation}.

	\begin{table*}[!htbp]
		
		\renewcommand{\arraystretch}{1.5}
		\caption{LAYERS AND PROPERTIES OF OUR NEURAL NETWORK}
		\label{table_example2}
		\centering
		
		\begin{tabular}{|l|l|l|l|l|l|l|l|}
			\hline
			\rowcolor{lightgray}
			\textbf{Dataset A} & \textbf{Input}& \textbf{Layer 1} &\textbf{Layer 2}&\textbf{Layer 3}&\textbf{Layer 4}& \multicolumn{2}{c|}{{\textbf{Properties}}} \\ 
			\hline
			{{No. Neurons}} & {12}  & {32} & {32} & {32} & {1} & {Learning rate / Optimizer} & {0.01 / Adam} \\
			\hline
			
			{{Activation f.}}	& {-}   & { Leaky Relu}& { Leaky Relu }& { Leaky Relu}& { Leaky Relu}& {No. epochs / Batch size} & {100 / 10}\\
			\hline
			
			{Regularisation}& {-} &{L2 ($\beta$=0.01)}& {L2 ($\beta$=0.01)}& {L2 ($\beta$=0.01)}& {L2 ($\beta$=0.01)}& {No. Samples (train / test)} &{57 / 23}\\
			\hline  
		\rowcolor{lightgray}
			\textbf{Dataset B} & \textbf{Input}& \textbf{Layer 1} &\textbf{Layer 2}&\textbf{Layer 3}&\textbf{Layer 4}& \multicolumn{2}{c|}{{\textbf{Properties}}} \\ 
			\hline
			
			{{No. Neurons}} & {12}  & {32} & {32} & {32} & {1} & {Learning rate / Optimizer} & {0.03 / Adam} \\
			\hline
			
			{{Activation f.}}	& {-}   & { Leaky Relu}& { Leaky Relu }& { Leaky Relu}& { Leaky Relu}& {No. epochs / Batch size} & {100 / 40}\\
			\hline
			
			{Regularisation}& {-} &{L2 ($\beta$=0.01)}& {L2 ($\beta$=0.01)}& {L2 ($\beta$=0.01)}& {L2 ($\beta$=0.01)}& {No. Samples (train / test)} &{224 / 23}\\
			\hline  
			
		\end{tabular}
	\end{table*}

	To tackle this issue, we need to normalize our data value in the range of 0 to 1. Several frameworks such as Pytorch \cite{paszke2019pytorch} and Tensorflow \cite{abadi2016tensorflow} are available and provide a competitive arsenal of tools to perform this operation. We have used the Pytorch framework in this experiment. ARTIST2 Language for WCET Flow Analysis (ALF) format is suitable for our approach because it contains both high-level and low-level constructs. Statement such as CALL and RETURN represent high-level constructs, and statements such as LOAD, STORE, and JUMP represent low-level constructs. Two datasets, A and B, are created. The total number of elements in the training datasets A and B are 57 and 224 respectively. 23 TacleBench programs have been used as testing data, and these programs are the same as those used in \cite{altenbernd2016early}. The training data is further divided into two parts: training, and validation through 5-fold cross-validation to check how well our model performs on the training data. The testing data is taken from the TacleBench benchmark which contains a test case for WCET analysis for different platforms. RMSE scores are used to compare our predicted value to the actual value. RMSE is calculated as the root of the mean of the squared differences between the predictions and the actual values.
	
	Gem5 \cite{binkert2011gem5} simulator is used to carry out all experiments to configure one processor with different attributes. The ARM810 processor is used with 5 stage pipeline, 8KB unified cache, MMU, and static branch prediction. Operations like floating points are implemented in the software. LLVM \cite{lattner2004llvm} compiler is used for compiling the test programs. Clang is used as a front-end to convert C source code into LLVM intermediate format. The LLVM Intermediate Representation (IR) file is given to ALF backend (C2ALF) to convert it into an ALF format, and the ARM backend (LLC) to convert it into an ARM object file. The SWEET tool is used to read the ALF code; it counts the number of different ALF constructs that appear as statements and operations. 	
	
	We use hyperparameter tuning \cite{feurer2019hyperparameter}, such as grid search and randomized search, to find the best possible neural network configuration by modifying the hyperparameters like learning rate (lr), number of epochs, and different optimization and activation functions in other experiments. The best structure of the deep learning model is shown in table \ref{table_example2}. We executed 12 different experiments by varying all the hyperparameters to find out the model that gives the least error on training and validation data. The comparison of different training and testing loss is shown in fig. 4. The X-axis represents a different experiment with hyperparameter tuning and the Y-axis loss values. In Figure 4(a)., the experiment with lr = 0.01 and neurons = 32 is the best one for dataset A as it gives the least error on both training and validation data. 
	The training loss error is significantly less than the validation loss error because dataset A has a considerably smaller sample program. Similarly, the experiment with lr = 0.03 and neurons = 32 is the best for dataset B. Learning curve of the best model for both datasets is shown in Figure 5. The loss scale is different in both figures because of the different numbers of the samples in each dataset. The figure indicates some oscillation in loss values at the beginning for dataset A and is smooth for dataset B, but as the number of epochs increases with time, loss values start converging and saturate to zero. Hence, we limit the number of training epochs to 100.

		\begin{figure*}[!t]
		\subfloat[Dataset A]{\includegraphics[width=3.5in]{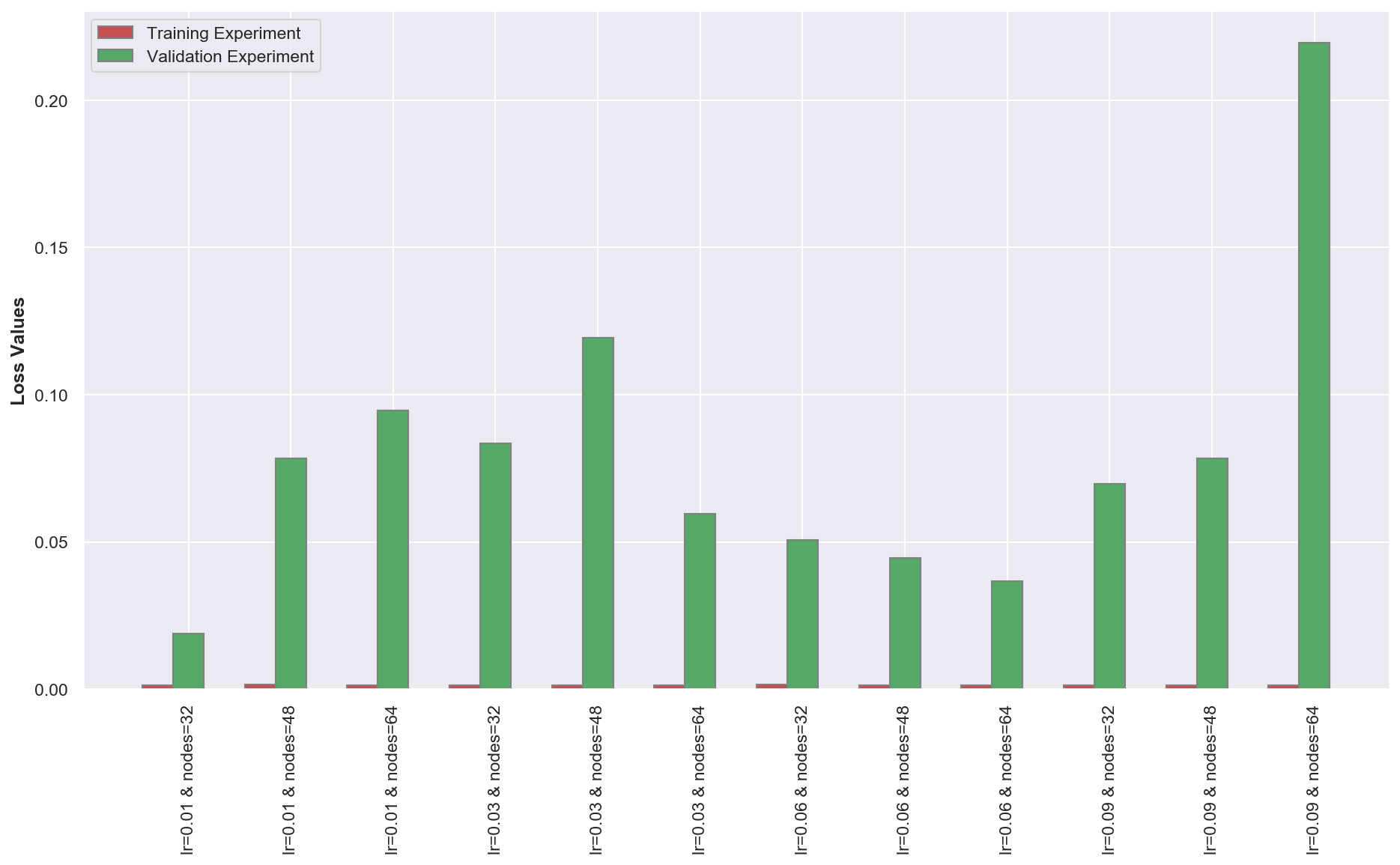}
			\label{fig_first_case1}}
		\hfill
		\subfloat[Dataset B]{\includegraphics[width=3.5in]{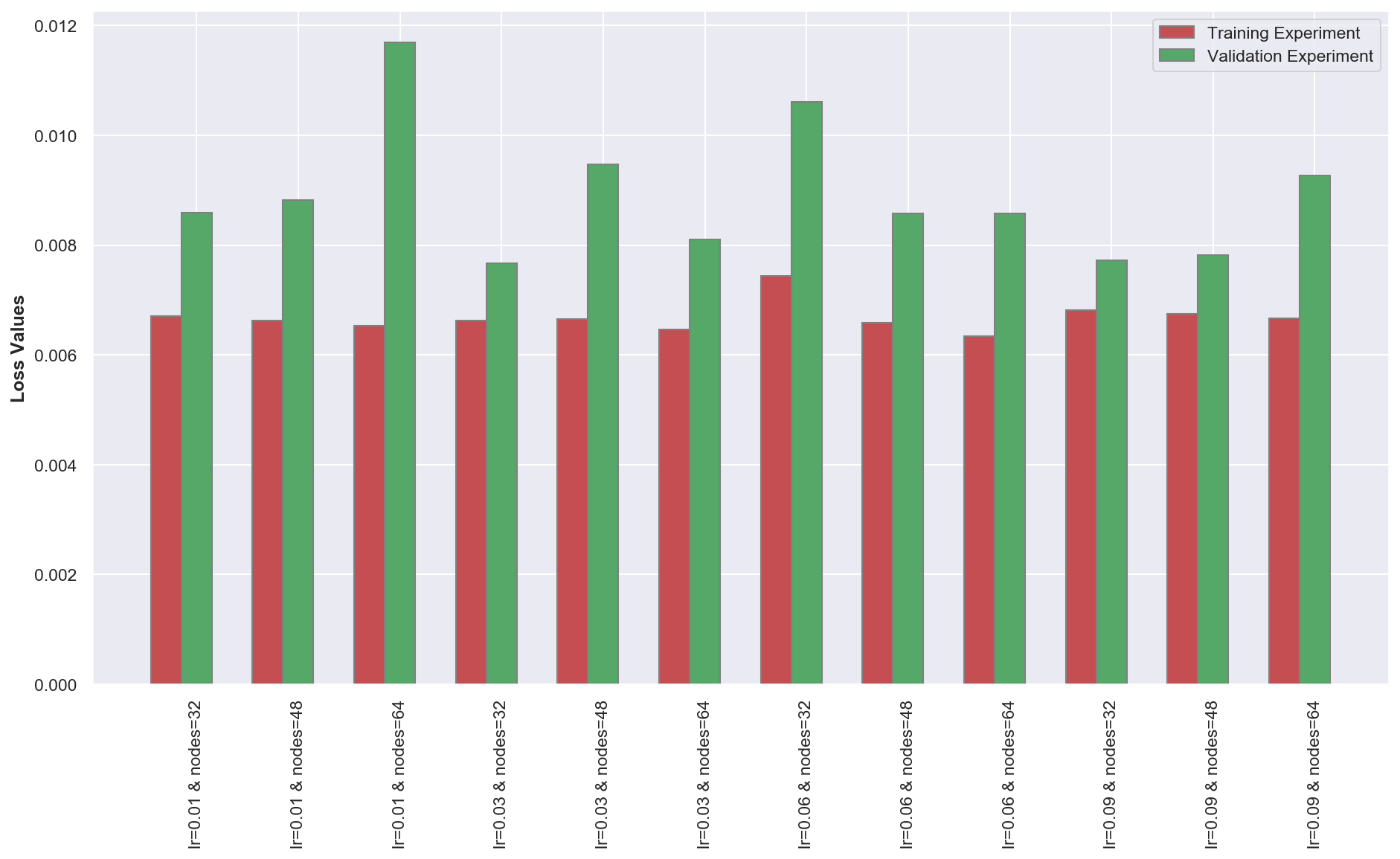}
			\label{fig_second_case2}}
		\caption{Comparison of Loss values using different configuration.}
		\label{fig_sim4}
	\end{figure*}
	
	\begin{figure*}[!t]
		\subfloat[Dataset A]{\includegraphics[width=3.5in]{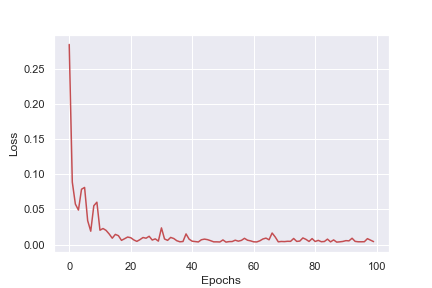}
			\label{fig_first_case3}}
		\hfill
		\subfloat[Dataset B]{\includegraphics[width=3.5in]{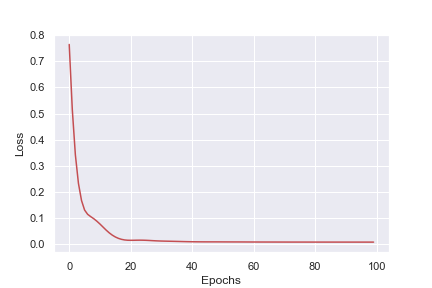}
			\label{fig_second_case4}}
		\caption{Comparison of Learning Curve.}
		\label{fig_sim3}
	\end{figure*}
	
	\begin{figure*}[!t]
		\subfloat[Dataset A]{\includegraphics[width=3.5in]{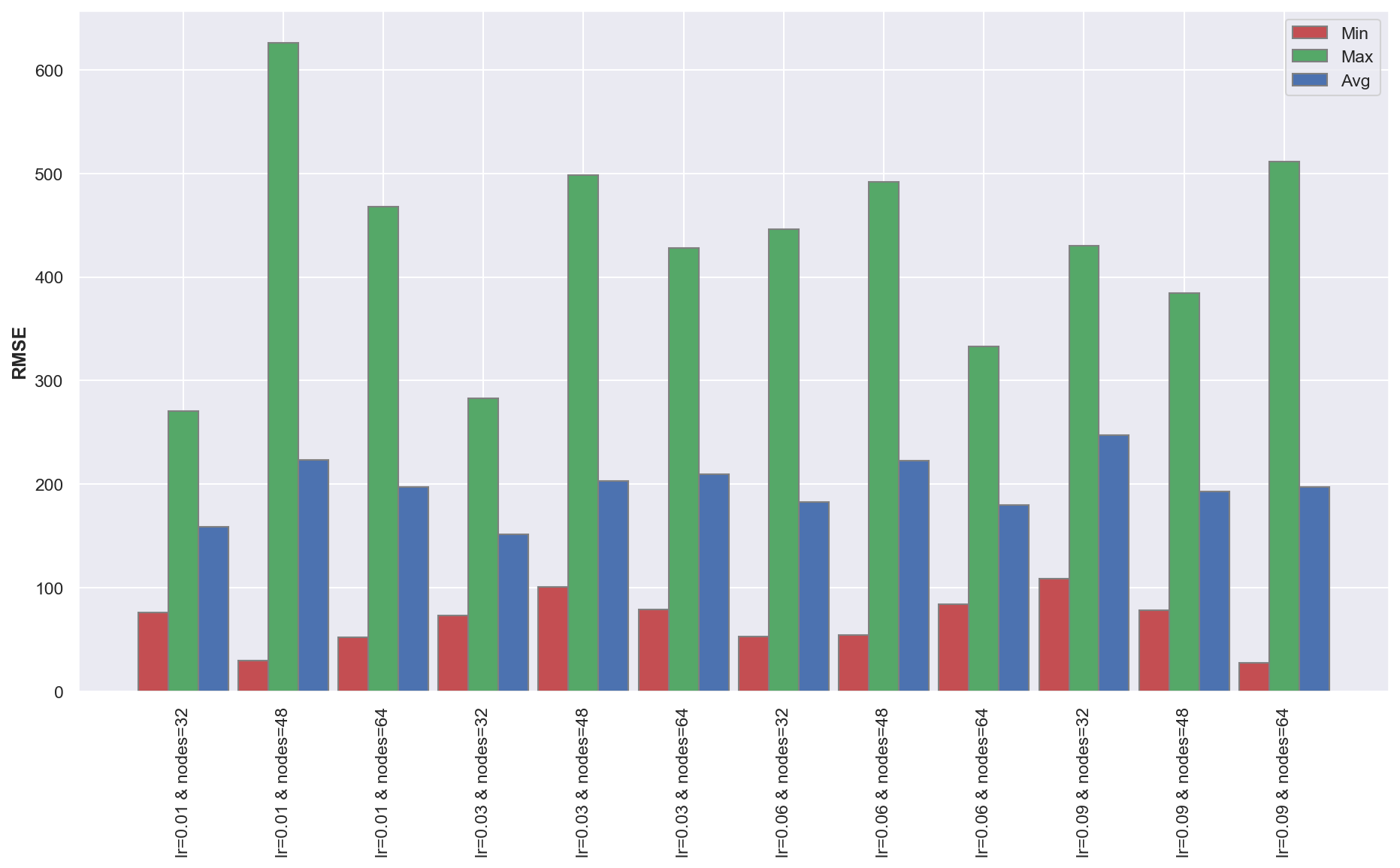}
			\label{fig_first_case5}}
		\hfill
		\subfloat[Dataset B]{\includegraphics[width=3.5in]{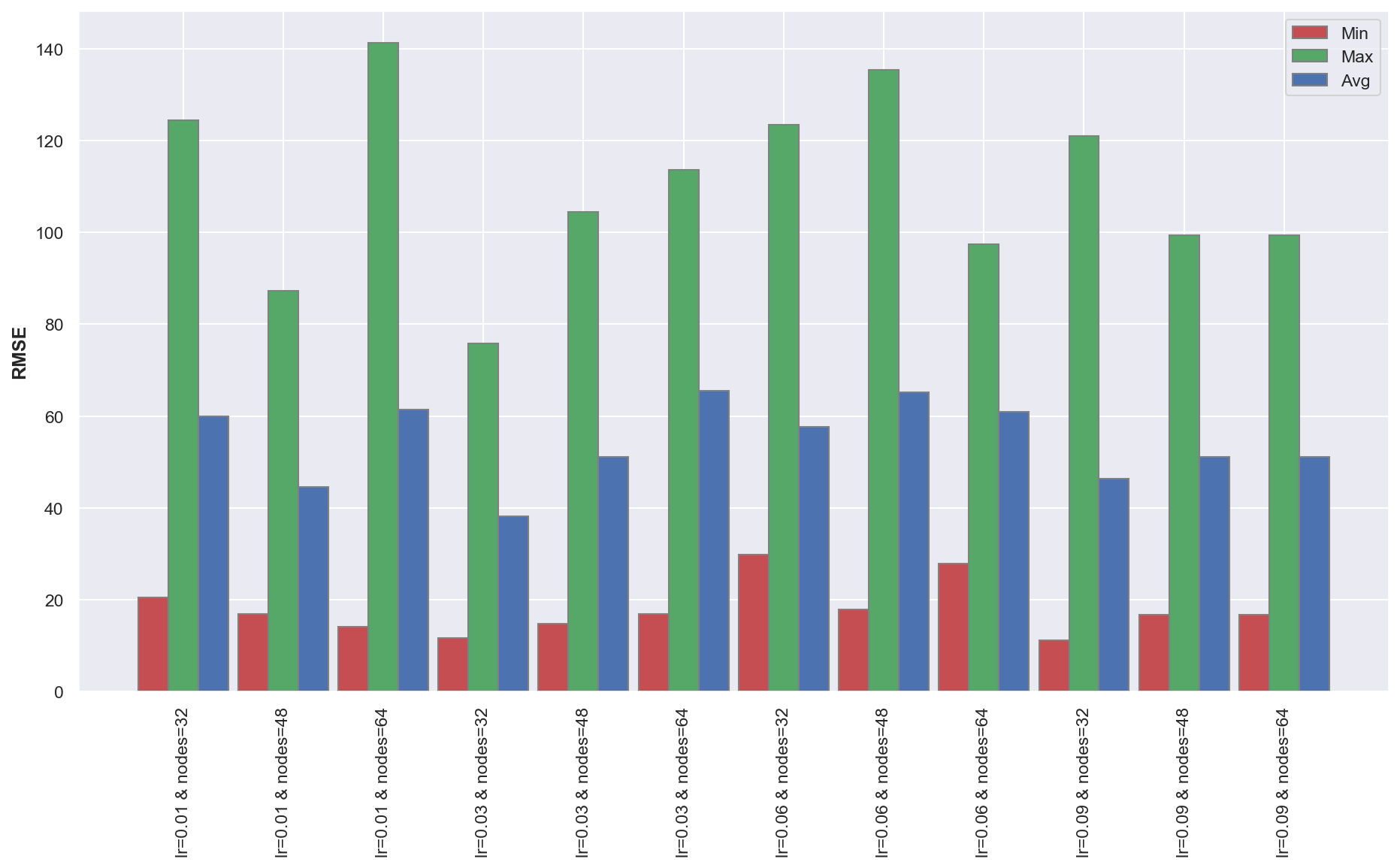}
			\label{fig_second_case6}}
		\caption{Comparison of Min, Max, and Avg RMSE values by executing the model using 12 different configuration settings.}
		\label{fig_sim2}
	\end{figure*}
	
		\begin{table*}[!h]
		\caption{RMSE ERRORS OF NEURAL NETWORK MODELS.}
		\begin{center}
			\begin{tabular}{lccccc}
				\hline
				\multirow{2}{*}{\textbf{Dataset}} & \multirow{2}{*}{\textbf{Average Error (RMSE)}}& \multirow{2}{*}{\textbf{Min. Error (RMSE)}} & \multirow{2}{*}{\textbf{Max. Error (RMSE)}}\\ \\
				& \textbf{Training / Test}&  \textbf{Training / Test} & \textbf{Training / Test}& \\
				\hline
				\multirow{2}{*}{\textbf{A}}& \multirow{2}{*}{{21\% / 41.3\%}}& \multirow{2}{*}{{17.8\% / 23\%}} & \multirow{2}{*}{{22.9\% / 66.7\%}} & \\ \\
				\hline
				\multirow{2}{*}{\textbf{B}}& \multirow{2}{*}{{12.7\% / 20.6\%}}& \multirow{2}{*}{{11.8\% / 17.4\%}}&  \multirow{2}{*}{{14.8\% / 23.5\%}}& \\ \\
				\hline
			\end{tabular}
			\label{tab2}
		\end{center}
	\end{table*}
  \section{{Results}}
  The results are shown in the graph in Figure 6; the percentage RMSE in table III  provide better insights into the model.
  We have executed our Deep learning model 12 times with each combination of different hyperparameters values. This allows us to calculate the minimum, maximum, and average error values for each configuration set up. The minimum, maximum, and average values are shown in Figure 6. For dataset A, we notice that the results with \textit{lr = 0.01} and \textit{nodes = 32} have shown better results where the minimum and average RMSE values are very close. Also, the maximum RMSE is comparatively close to these values. The trend in dataset B is different as the results with \textit{lr = 0.03} and \textit{nodes = 32} have shown the lowest error rate. Although there is variation in results, there is not a big difference between the minimum and maximum error values, which shows that, in most cases, our model has lower error rates  as compared to few cases where the error rate is high. The large error is due to the considerable difference between the properties of training and testing datasets. 

    We compare our method with other methods in the literature by also pointing out similarities and differences between the two:
    \begin{itemize}
      \item We use statements and operations of source programs as a feature, similar to the approaches presented in \citep{bonenfant2017early,huybrechts2018new,oyamada2004accurate,huybrechts2018introduction}.
      \item Unlike \cite{bonenfant2017early} and \cite{huybrechts2018new},  we do not use a static approach.  Measurement-based approach is used in a target-hardware agnostic manner.
      \item Unlike \cite{oyamada2004accurate}  we do not use assembly instructions for feature categorization. We have used a source program to extract the features.
      \item Like in \cite{huybrechts2018introduction}  we have used Root Mean Square Error (RMSE). However, our results are different because we have used different datasets and WCET strategies. \\ \\ \\ \\ \\ \\ \\ 
    \end{itemize}

  \section{{Conclusions and Future Work}}
  This paper presents an approach that can be used to predict early WCET using a Deep Neural Network. The model estimates WCET from the source code of the applications. Features are generated using the SWEET tool, i.e., the number of statements and operations in the source code, which are fed into our networks to predict WCET after scaling data. Two datasets, A and B, are created with 57 and 224 samples respectively. We demonstrate the model for ARM processors. We have used the Pytorch framework to implement a feed-forward neural network that converges quickly. The model performance is evaluated using a metric called the RMSE. We calculated the minimum, maximum, and average RMSE for each distinct neural network configuration. The RMSE value for the bigger dataset is much better than the smaller dataset. The results shown in this paper are not promising as it is too large to use as an upper bound. However, getting these numbers in the early stage of developing a system is useful in many ways, such as preventing systems designers from assuming a pessimistic upper bound on the WCET.

  We believe that Deep Neural Networks can be applied to improve these results further. Instead of using a simulator for measurement, we can use a real processor to obtain more accurate models. This is a subject for future research, too. We intend to train deeper models on bigger datasets to obtain even better and reliable results in the future. Deep learning has good potential in WCET analysis. Additionally, we can further reduce WCET analysis results' pessimism by applying different models like Convolution Neural Network (CNN) and Recurrent Neural Network (RNN) that can be used to estimate early WCET.

  \section*{Acknowledgment}
  The authors would like to thank Sourav Mishra for his valuable suggestions on several aspects of this paper. We would also like to thank the anonymous reviewers for their insightful comments and feedback.


  \bibliographystyle{IEEEtran}
  \bibliography{sample}

\end{document}